\documentclass[sigconf,screen]{acmart}
\usepackage{amsmath,amsfonts,bm}

\def\eqref#1{equation~\ref{#1}}
\def\1{\bm{1}}

\def\rr{{\textnormal{r}}}

\def\rvr{{\mathbf{r}}}

\def\ervr{{\textnormal{r}}}

\def\vb{{\bm{b}}}

\def\vs{{\bm{s}}}

\def\vy{{\bm{y}}}

\def\evs{{s}}

\def\mW{{\bm{W}}}
\def\mX{{\bm{X}}}
\def\mY{{\bm{Y}}}

\DeclareMathAlphabet{\mathsfit}{\encodingdefault}{\sfdefault}{m}{sl}
\SetMathAlphabet{\mathsfit}{bold}{\encodingdefault}{\sfdefault}{bx}{n}

\usepackage{multirow}
\usepackage{hyperref}
\usepackage{url}
\usepackage{subfig,graphicx}
\usepackage{multicol}
\usepackage{booktabs}
\usepackage{pifont}
\usepackage{bbding}
\usepackage{enumitem}
\usepackage{wrapfig}
\usepackage{svg}
\usepackage{listings}
\usepackage{xcolor}
\usepackage{makecell}
\usepackage{cleveref}
\usepackage{enumitem}
\usepackage{CJK}

\usepackage{tcolorbox}
\def\BibTeX{{\rm B\kern-.05em{\sc i\kern-.025em b}\kern-.08em
    T\kern-.1667em\lower.7ex\hbox{E}\kern-.125emX}}

\usepackage{xspace}
\newcommand{\toolname}{\textit{ModelObfuscator}\xspace}
\newcommand{\thirdObfuscateStrategyName}{neural structure obfuscation\xspace}
\newcommand{\ThirdObfuscateStrategyName}{Neural structure obfuscation\xspace}

\makeatletter
\DeclareRobustCommand\bmvaOneDots{\futurelet\@let@token\bmv@onedotsaux}
\def\bmv@onedotsaux{\ifx\@let@token.,\else.,\null\fi\xspace}
\DeclareRobustCommand\bmvaOneDot{\futurelet\@let@token\bmv@onedotaux}
\def\bmv@onedotaux{\ifx\@let@token.\else.\null\fi\xspace}
\makeatother
\def\eg{\emph{e.g}\bmvaOneDots} 
\def\ie{\emph{i.e}\bmvaOneDots} 
 
\def\etc{\emph{etc}.} 
\def\vs{\emph{vs}\bmvaOneDot}
 
\def\etal{\emph{et al}\bmvaOneDot}

\setcopyright{acmlicensed}
\acmPrice{15.00}
\acmDOI{10.1145/3597926.3598113}
\acmYear{2023}
\copyrightyear{2023}
\acmSubmissionID{issta23main-p275-p}
\acmISBN{979-8-4007-0221-1/23/07}
\acmConference[ISSTA '23]{Proceedings of the 32nd ACM SIGSOFT International Symposium on Software Testing and Analysis}{July 17--21, 2023}{Seattle, WA, USA}
\acmBooktitle{Proceedings of the 32nd ACM SIGSOFT International Symposium on Software Testing and Analysis (ISSTA '23), July 17--21, 2023, Seattle, WA, USA}
\received{2023-02-16}
\received[accepted]{2023-05-03}

\begin{document}

%%
%% The "title" command has an optional parameter,
%% allowing the author to define a "short title" to be used in page headers.
\title{\toolname: Obfuscating Model Information to Protect Deployed ML-Based Systems}

\author{Mingyi Zhou}
\email{mingyi.zhou@monash.edu}
\affiliation{%
  \institution{Monash University}
  \city{Melbourne}
  \state{VIC}
  \country{Australia}
}
\author{Xiang Gao}
\email{xiang_gao@buaa.edu.cn}
\affiliation{%
  \institution{Beihang University}
  \city{Beijing}
  \country{China}
}
\author{Jing Wu}
\email{jing.wu1@monash.edu}
\affiliation{%
  \institution{Monash University}
  \city{Melbourne}
  \state{VIC}
  \country{Australia}
}
\author{John Grundy} 
\email{john.grundy@monash.edu}
\affiliation{%
  \institution{Monash University}
  \city{Melbourne}
  \state{VIC}
  \country{Australia}
}
\author{Xiao Chen}\authornote{Dr. Li Li was a senior lecturer at Monash. He supervised this project for the whole period. Corresponding authors: Xiao Chen and Li Li.}
\email{xiao.chen@monash.edu}
\affiliation{%
  \institution{Monash University}
  \city{Melbourne}
  \state{VIC}
  \country{Australia}
}
\author{Chunyang Chen}
\email{chunyang.chen@monash.edu}
\affiliation{%
  \institution{Monash University}
  \city{Melbourne}
  \state{VIC}
  \country{Australia}
}
\author{Li Li}\authornotemark[1]
\email{lilicoding@ieee.org}
\affiliation{%
  \institution{Beihang University}
  \city{Beijing}
  \country{China}
}

\begin{abstract}
More and more edge devices and mobile apps are leveraging deep learning (DL) capabilities.
Deploying such models on devices -- referred to as on-device models -- rather than as remote cloud-hosted services, has gained popularity because it avoids transmitting user's data off of the device and achieves high response time.
However, on-device models can be easily attacked, as they can be accessed by unpacking corresponding apps and the model is fully exposed to attackers.
Recent studies show that attackers can easily generate white-box-like attacks for an on-device model or even inverse its training data.
To protect on-device models from white-box attacks, we propose a novel technique called \textit{model obfuscation}.
Specifically, model obfuscation hides and obfuscates the key information -- structure, parameters and attributes -- of models by \textit{renaming}, \textit{parameter encapsulation}, \textit{\thirdObfuscateStrategyName obfuscation}, \textit{shortcut injection}, and \textit{extra layer injection}.
We have developed a prototype tool \toolname to automatically obfuscate on-device TFLite models.
Our experiments show that this proposed approach can dramatically improve model security by significantly increasing the difficulty of parsing models' inner information, without increasing the latency of DL models.
Our proposed on-device model obfuscation has the potential to be a fundamental technique for on-device model deployment.
Our prototype tool is publicly available at \url{https://github.com/zhoumingyi/ModelObfuscator}.
\end{abstract}

\begin{CCSXML}
<ccs2012>
   <concept>
       <concept_id>10011007.10010940.10011003.10011114</concept_id>
       <concept_desc>Software and its engineering~Software safety</concept_desc>
       <concept_significance>500</concept_significance>
       </concept>
   <concept>
       <concept_id>10011007.10010940.10011003.10011004</concept_id>
       <concept_desc>Software and its engineering~Software reliability</concept_desc>
       <concept_significance>500</concept_significance>
       </concept>
 </ccs2012>
\end{CCSXML}

\ccsdesc[500]{Software and its engineering~Software safety}
\ccsdesc[500]{Software and its engineering~Software reliability}

\keywords{SE for AI, AI safety, model obfuscation, model deployment}

\maketitle

\section{Introduction}
This study focuses on how to defend against software analysis and reverse engineering that collects the inner information of on-device models. These kinds of attacks are far more effective than model-stealing techniques such as substitute model training or side-channel attacks, since on-device models are directly hosted on mobile devices. 

Numerous edge and mobile devices are leveraging deep learning (DL) capabilities.
Though DL models can be deployed on a cloud platform, data transmission between mobile devices and the cloud may compromise user privacy and suffer from severe latency and throughput issues. 
To achieve high-level security, users' personal data should not be sent outside the device. To achieve high throughput and short response time, especially for a large number of devices, on-device DL models are needed.
The capabilities of newer mobile devices and edge devices keep increasing, with more powerful systems on a chip (SoCs)
and a large amount of memory, making them suitable for running on-device models. 
Indeed, many intelligent applications have already been deployed on devices~\citep{xu2019first} and benefited millions of users. 

\begin{figure}[ht]
  \begin{center}
    \includegraphics[width=0.85\linewidth]{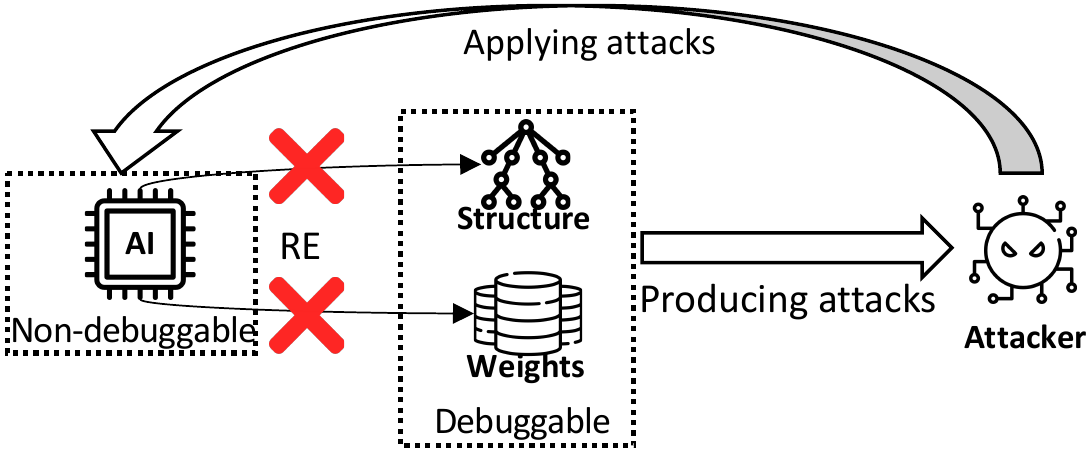}
  \end{center}
  % \vspace{-1em}
  \caption{The typical scenarios of attacking on-device models. Our approach can disable the reverse engineering (\ie RE).}
  \label{fig:attack_motivation}
\end{figure}

Unfortunately, since on-device models are directly hosted on mobile devices, adversaries can easily unpack the mobile apps to locate the models for security exploitation. Such on-device models are thus facing more serious security threats. To protect the on-device model, on-device DL frameworks like TFLite are released to users as black-box ones, \ie, the released TFLite models do not support the gradient computing.
Since gradient information is considered crucial to implement effective white-box attacks,
preventing attackers from obtaining gradient information from on-device models (referred to as non-debuggable models) could enhance model's security.
However, it does not fulfill such a purpose in practice, which is shown in Figure~\ref{fig:attack_motivation}. Attackers can parse the information of models (\eg model structures and weights), generate efficient attacks based on the parsed information, and apply the attacks to the target on-device model.
For example, Huang \etal~\cite{huang2021robustness} parse features of the on-device model to find a surrogate model from the web, which can then be used to launch transferable adversarial attacks on mobile models. Li \etal~\cite{li2021deeppayload} developed a method to perform backdoor attacks on mobile models. This method uses reverse engineering to parse the model file and inject a triggering model into the model file. As the attackers can easily obtain the explicit information of on-device models through public APIs or reverse engineering, on-device models are at serious risk that attackers can also perform many other attacks for on-device models, such as model inversion attacks, membership inference attacks, \etc~\citep{szegedy2013intriguing,chen2017targeted,shokri2017membership,fang2020local}.  
However, to the best of our knowledge, no effective defense method has been proposed to resist reverse engineering for on-device DL models. 

Borrowing the idea of code obfuscation, which is a well-developed approach for hiding key information in software, we propose to
obfuscate the information of on-device ML models. As the on-device model interacts with the DL libraries like TFLite at runtime, we also need to obfuscate the information of model files (\eg .tflite files) and modify the DL libraries to execute the correct forward computation of original models.
In this paper, \textbf{we propose a novel on-device ML model protection approach based on model obfuscation, which focuses on improving AI safety for disabling the model parsing using software analysis or reverse engineering}.
Given a trained model and its underlying DL library (\eg TFLite),
an end2end prototype tool, \toolname, is developed to generate obfuscated on-device model and corresponding DL library. 
\toolname first extracts the information of the target model and locates the source code in the library used by its layers.
It then obfuscates the information of the models and builds a customized DL library that is compatible with the obfuscated model.
To achieve this, we design five obfuscation methods, including: (1) \textit{renaming}, (2) \textit{parameter encapsulation}, (3) \textit{\thirdObfuscateStrategyName}, (4) \textit{random shortcut injection}, and (5) \textit{random extra layer injection}. 
These obfuscation methods significantly increase the difficulty of parsing the information of the model. 
Our on-device ML model obfuscation can prevent attackers from reconstructing the model.
It is also hard for attackers to transfer the trained weights and structure of models to steal intellectual property using model conversion, because the connection between the obfuscated information and the original one is randomly generated.
Experiments on 10 different models show that \toolname can against state-of-the-art model parsing and attack tools with a negligible time overhead and 20\% storage overhead. The codes of \toolname are shared at \url{https://github.com/zhoumingyi/ModelObfuscator}.

The key contributions in this work include: 
\begin{itemize}[leftmargin=*]
    \item We propose a novel model obfuscation framework to hide the key information of deployed DL models to resist model parsing. It can prevent attackers from generating efficient attacks and stealing the knowledge of on-device models by significantly increasing the cost of attacks.
    \item We design five obfuscation strategies for protecting on-device models and implement an end2end prototype tool, \toolname. This tool automatically obfuscates the model and builds a compatible DL software library. The tool is publicly available.
    \item We provide a taxonomy and comparison of different obfuscation methods in terms of effectiveness and overhead, to guide model owners in choosing appropriate defence strategies.
\end{itemize}

\section{Background and Related Work}

\subsection{On-Device DL Models}

\paragraph{\textbf{DL frameworks}}
The open-source community has developed many well-known open-source frameworks for DL tasks such as TensorFlow~\cite{tensorflow2015_whitepaper}, Theano~\cite{al2016theano}, Caffe~\cite{jia2014caffe}, Keras~\cite{chollet2018keras}, and PyTorch~\cite{paszke2019pytorch}. These frameworks dominate the development of DL models and set standards for them~\cite{dilhara2021understanding}. PyTorch is one of the latest DL frameworks, and is gaining popularity for its ease of use and its capability to construct the dynamic computational graph, which is now widely used by the academic community. In contrast, TensorFlow is widely used by companies, startups, and business firms to automate things and develop new systems. It has distributed training support, scalable production options, and support for mobile devices. Currently, companies and research teams have made huge efforts to develop open-source on-device frameworks like TensorFlow Lite (TFLite), Caffe2, Caffe, NCNN, and ONNX. As an on-device DL platform, TensorFlow Lite (TFLite) is the most popular framework for DL models on smartphones, as it has GPU support and is optimized for mobile devices~\cite{xu2019first,huang2021robustness}. 

\paragraph{\textbf{Deployed DL models}}
As the training of DL models is intensive in both data and computing, mobile developers often collect the data and train their models on the computing server prior to app deployment. 
Developers also need to specifically compile the trained models to be compatible with devices (\ie on-device models) so as to speed up the model inference on mobile CPU/GPUs~\cite{chen2021empirical,chen2020comprehensive,niu2021dnnfusion}.
At app installation time, the trained models are deployed, along with the app code itself, in the installation package of an app. At runtime, apps perform the inference of DL models by invoking application programming interfaces (APIs) of DL frameworks, and subsequently achieve AI capabilities.
% such as image classification and text recognition.

\subsection{TFLite Models}
TFLite models have powerful features for running models on edge devices but it does not provide APIs to access the gradient or intermediate outputs like other TensorFlow or PyTorch models. TensorFlow provides a \textit{TensorFlow Lite Converter}\footnote{\href{https://www.tensorflow.org/lite/convert/index}{https://www.tensorflow.org/lite/convert/index}} to convert a TensorFlow model into a TensorFlow Lite model. 
TFLite models run on the FlatBuffers Platform\footnote{\href{https://google.github.io/flatbuffers/}{https://google.github.io/flatbuffers/}}. FlatBuffers is an efficient cross-platform serialization library for multiple programming languages. It has several advantages for employing the model on mobile devices that it can access serialized data without parsing/unpacking and only needs small computational resources. TFLite uses a schema file to define the data structures. For parsing the model structure and weights from the \texttt{.tflite} file, we can use the schema file \footnote{\href{https://github.com/tensorflow/tensorflow/blob/master/tensorflow/lite/schema/schema.fbs}{schema file (The link is too long to display)}} of TFLite to parse FlatBuffers and get the JSON file that contains detailed information of the \texttt{.tflite} file. 
% \vspace{-0.8em}

\begin{figure*}[tb]
% \vspace{-1em}
  % \vspace{-6pt}
  \begin{center}
    \includegraphics[width=0.85\textwidth]{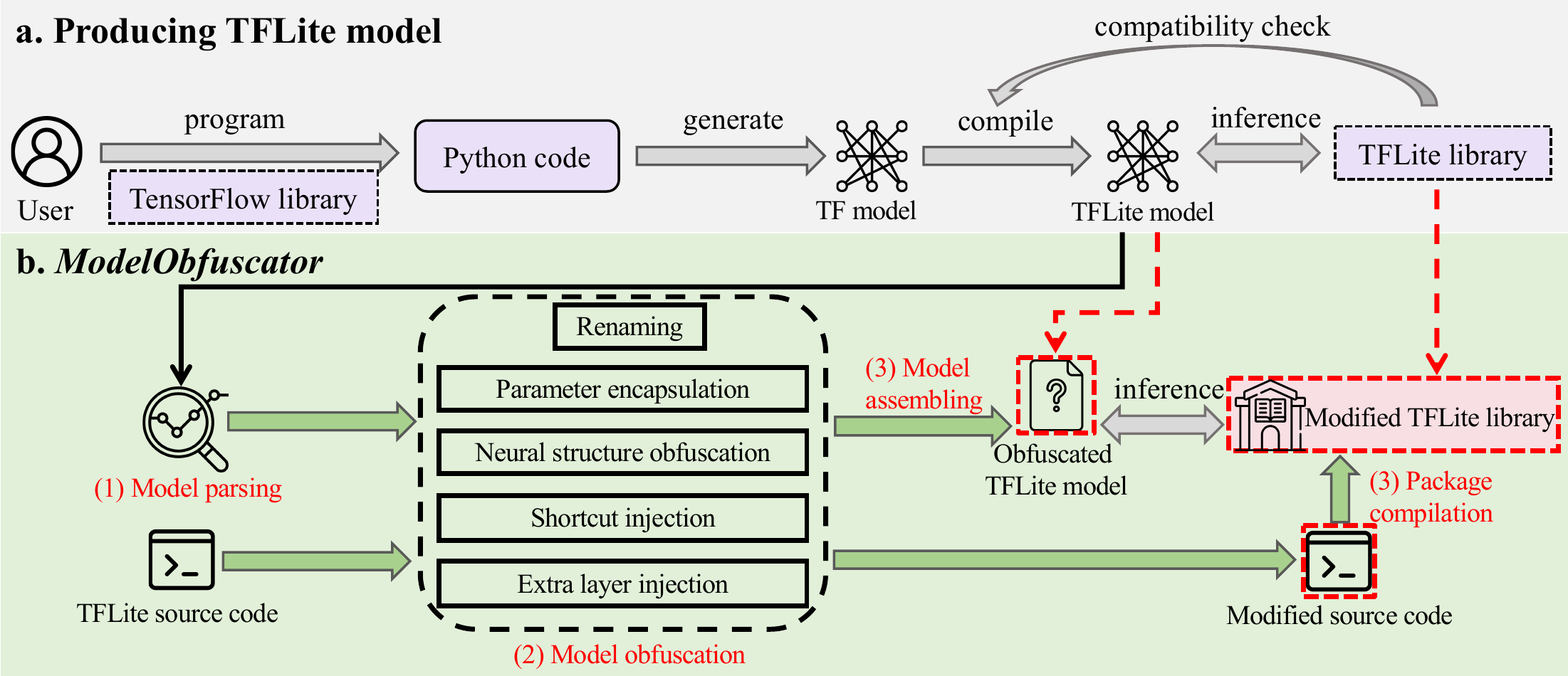}
  \end{center}
  % \vspace{-6pt}
  \caption{The working process of our \toolname on-device DL model obfuscation tool. `a': the process of producing TFLite model. `b': the framework of the proposed \toolname.}
  \label{fig:workflow}
  % \vspace{-1em}
\end{figure*}

\subsection{Existing Model Parsing and Defenses}
The key information in a deployed DL model can be parsed by three kinds of approaches: (1) Parsing the model's weights through queries~\citep{tramer2016stealing,zhou2020dast}. Attackers use samples to query the outputs of the target model. Then, they can train a similar model with target model through the samples and outputs. (2) Parsing the model's architecture by side-channel attacks\citep{hu2020deepsniffer,wei2020leaky,li2021neurobfuscator}. Attackers can estimate the architecture of the neural networks by the side channel information (\eg latency and DRAM accesses). (3) Parsing the entire model from devices using software analysis or reverse engineering~\citep{vallee2010soot,li2021deeppayload}. 
Although many attacks have been proposed to extract DL models, it is hard for adversaries to precisely reconstruct DL models that are identical to the original ones using queries or side-channel information. 
These attacks cannot access the inner information of the model, meaning that they are black-box attacks.
Among these three approaches, reconstructing DL models using queries or side-channel information is hard to obtain identical models to the original ones.
Since, the first two approaches cannot access the inner information of models, they only enable black-box attacks.
In contrast, since on-device models are delivered in mobile apps and hosted on mobile devices, adversaries can easily unpack the mobile apps to parse the original models for security exploitation, which enables white-box attacks.
As shown in previous work, white-box attacks can achieve a much higher success rate than black-box attacks~\cite{zhang2022investigating}.
Apart from being attacked, deploying DL models on users' devices may cause intellectual property leakage~\citep{sun2021mind}.

Existing model defense methods 
%to resist the model extraction attacks 
can be categorized into two different groups: (1) defend against query-based attacks, and (2) defend against side-channel attacks.
For defending against query-based attacks, some studies~\citep{orekondy2019prediction,kariyappa2020defending,mazeika2022steer,xu2018deepobfuscation,szentannai2019mimosanet,szentannai2020preventing} propose different strategies to degenerate the effectiveness of query-based model extraction.
While other studies~\citep{xu2018deepobfuscation,szentannai2019mimosanet,szentannai2020preventing} propose methods to train a simulating model, which has similar performance to the original model, but is more resilient to query-based attacks.
For securing the AI model at the side-channel level, a recent work modifies the CPU and Memory costs to resist the model extraction attacks~\citep{li2021neurobfuscator}. 
However, existing defenses have not yet been aware of model parsing using software analysis or reverse engineering, which is more harmful than other kinds of model parsing for deployed DL models. Our method is proposed to defend against the parsing using software analysis or reverse engineering. 

\subsection{Reverse Engineering Tools for DL Models}
We want an approach that is highly resistant to model parsing that uses software analysis or reverse engineering tools.  Existing model parsing methods using reversing engineering fall into three categories: (1) model format conversion -- existing model conversion tools access the structure and weights of non-debuggable on-device models through public APIs, and then assemble them into a new model with a different format. 
(2) model parsing in the buffer -- the on-device model format TFLite loads the data on FlatBuffer~\footnote{\url{https://google.github.io/flatbuffers/}}. Attackers can then extract the model structure and weights in FlatBuffer~\cite{li2021deeppayload}. 
(3) finding a similar differentiable model from the Internet by comparing the features among models -- the current mainstream on-device model format TFLite does not support some advanced functions like auto-differentiation to protect the deployed model. But attackers like the App Attack can find a differentiable model with similar structure and weights from the Internet~\citep{huang2022smart}.

\subsection{Code Obfuscation} Code obfuscation methods are initially developed for hiding the functionality of the malware. Then, the software industry also uses it against reverse engineering~\citep{schrittwieser2016protecting}. They provide complex obfuscating algorithms for programs like JAVA code~\citep{collberg1997taxonomy,collberg1998manufacturing}, including robust methods for high-level languages~\citep{wang2001security} and machine code level~\citep{wroblewski2002general} obfuscation. Code obfuscation is a well-developed technique to secure the source code. However, traditional code obfuscation approaches are hard to protect on-device models, especially for protecting the structure of the models and their parameters. In this work, inspired by traditional code obfuscation, we propose a novel model obfuscation approach to obfuscate the model files and then produce a corresponding DL library for them.

\section{The \toolname Solution}
%\subsection{Overview}
%\vspace{-0.5em}
In this section, we will introduce the threat model in this study and show the detail of the proposed \toolname.
\subsection{ML Platform and Threat Model}

\paragraph{\textbf{On-device ML model platform}}
We chose the TensorFlow Lite (TFLite) platform to demonstrate our model obfuscation approach. TFLite is currently the most commonly used on-device DL platform.
The steps to produce TFLite models are shown in the top half of Figure~\ref{fig:workflow}.
Usually, DL developers use TensorFlow application programming interfaces (APIs) to define and train TensorFlow (TF) models.
%They then compile the TF model into a TFLite model.
The trained TF models are then compiled into a TFLite models.
Note that TFLite has different implementations with TF, and their operators (\ie layers) may not be compatible. 
Therefore, when compiling TFLite models, the TFLite library will check their compatibility. 
Once the compatibility check passes, the compiled TFLite model can run on devices using the TFLite library. 
% Hence, the problem of obfuscating TFLite models is to design obfuscation strategies and make the obfuscated model compatible with the TFLite library.

\paragraph{\textbf{Threat model}}
The on-device model is usually saved as a separate file (e.g., .tflite file) and packed into the app package. 
Attackers can either download the target app from the app markets (e.g., Google Play and iOS App store) or extract the app package file (e.g., APK file for Android, and IPA file for iOS) from the hosting devices. These app package files can then be decompiled by off-the-shelf reverse-engineering tools (e.g., Apktool~\footnote{\url{https://ibotpeaches.github.io/Apktool/}} and IDA Pro~\footnote{\url{https://hex-rays.com/ida-pro/}}) to get the original DL model file. 
Although many on-device DL platforms do not support some advanced functions like backpropagation, attackers can assemble the model architecture and weights into a differentiable model format~\cite{huang2021robustness}, or they can use software analysis methods to generate attacks for the target model ~\citep{li2021deeppayload,huang2022smart}. 
In this paper, we will obfuscate the information of the model and underlying DL library to prevent the software analysis and model conversion tools from getting model's key information. %and generate a compatible DL library for the obfuscated model that only supports the forward inference function.

We analyzed the current mainstream DL platforms (\ie TensorFlow and PyTorch) and identified the following two main observations.
First, these platforms are open-source and provide a set of tools to build the library (\eg TFLite library) from the source code.
Second, they officially support customized operators (\eg neural layers). Specifically, on top of these DL platforms, users could implement customized layers in C/C++ and compile customized layers to executable files (\texttt{.so} file in TensorFlow).
Then, users can use these customized layers via high-level Python interfaces.
%So, we can define an obfuscated layer, which does not have explicit or correct information of the original one, then locate the source code of the layer, and refer to the source code to implement the correct data flow in the customized layer.
% The customized layer will support the compilation and inference of obfuscated layers.
% As a third party, the model obfuscation framework shown in Figure~\ref{fig:workflow} can be applied to different DL platforms if they support the aforementioned two functions.
Those features enable us to design obfuscation techniques to obfuscate model and DL library code together.
The bottom half of Figure~\ref{fig:workflow} shows the overview of our model obfuscation framework.
Specifically, \toolname has three main steps: (1) model parsing, (2) model obfuscation, and (3) model assembling \& library recompilation.
In the following subsections, we detail the proposed \toolname.

\subsection{Model Parsing}

\begin{figure*}[tb]
% \vspace{-0.5em}
  \begin{center}
    \includegraphics[width=0.9\textwidth]{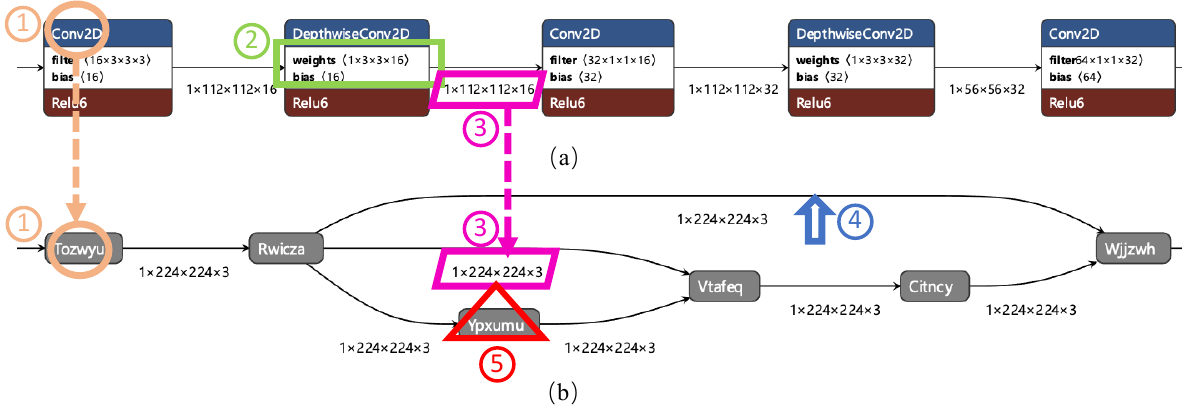}
  \end{center}
  \vspace{-1em}
  \caption{Example of five obfuscation strategies for the TFLite model obtained from a real-world fruit recognition app. (a) part of the original model. (b) the corresponding obfuscated model. `{\color{orange}\ding{192}}': \textit{renaming}. `{\color{green}\ding{193}}': \textit{parameter encapsulation}. `{\color{purple}\ding{194}}': \textit{neural structure obfuscation}. `{\color{blue}\ding{195}}': \textit{shortcut injection}. `{\color{red}\ding{196}}': \textit{extra layer injection}. This visualization is generated by Netron~\citep{roeder_lutz_2017_7109451}, which is a well-known visualization platform for neural networks.}
  \label{fig:injection}
  % \vspace{-1em}
\end{figure*}

The first step of \toolname model obfuscation is to parse the deployed model to extract its key information.
% Normally, each layer of the DL model may have multiple attributes (\eg stride and padding in the convolution layer), parameters, inputs, and outputs. 
\toolname first extracts the structure information of each layer, including the name of layers (\eg \texttt{Conv2D}) and model structures (including model's input, output, layer ID and \etc).
The extracted structure information is depicted using JSON format, which can be visualized as Figure~\ref{fig:injection} (a). It includes all neural layers and their spatial relationships.
%spatial relationships, to assemble a minimal model structure.
%On the right is an example of minimal model structure, which only contains the necessary elements to depict the model structure.
Then, \toolname extracts the parameter of each layer, such as the layer's configurations and weights.
Moreover, \toolname identifies the source code used by each layer by referring to the underlying libraries.
% the source code defines the function of each layer.
The identified source code includes relevant packages and functions of the TFLite layers.

\subsection{Model Obfuscation}
After extracting the on-device ML model information and its corresponding source codes, \toolname will obfuscate the model as well as the source codes.
\toolname uses five obfuscation strategies:  \textit{renaming}, \textit{parameter encapsulation}, \textit{\thirdObfuscateStrategyName}, \textit{shortcut injection}, and \textit{extra layer injection}. We describe the implementation details of each of these obfuscation strategies in the following subsections.

% \begin{minipage}[b]{0.64\linewidth}
% \vspace{-0.7em}
\paragraph{\textbf{\textit{Renaming}}} The most straightforward obfuscation strategy is the renaming of a layer.
Usually, the layer's name contains important information, which is the function of this layer.
For instance, ``Conv2DOptions'' indicates a 2D convolution layer. 
Such information is useful for attackers to reconstruct the model to generate white-box attacks or to obtain a similar surrogate model to conduct effective black-box attacks.
To hide such important information, we randomly modify each layer's name. The {\color{orange}\ding{192}} of Figure~\ref{fig:injection} is an example of an obfuscated \texttt{Conv2D} layer. 
\toolname automatically replaces the real name with the random meaningless string \texttt{Tozwyu}. Note that the same layers in the TFLite model will have different random names.
Meanwhile, \toolname creates a copy of \texttt{Conv2D}'s source code and replaces the layer name (\ie \texttt{Conv2D}) in the source code with \texttt{Tozwyu}.
Note that we modify the duplicate of \texttt{Conv2D} in case the modification affects other parts of the TFLite library.
After adding modified source codes to the TFLite project, the recompiled TFLite library will recognize obfuscated layers as custom layers and correctly execute them at runtime.

% \vspace{-0.7em}
\paragraph{\textbf{\textit{Parameter encapsulation}}} Existing TFLite models have two main assets: model structure and parameters. The parameter can be obtained in the training period. 
Given an input, a TFLite model will compute the results using the input tensor and parameters that are stored in the model file. 
As we discussed above, parameter exposure is very dangerous.
An adversary could use the parameter information to perform many kinds of white-box attacks, \eg adversarial attacks and model inversion attacks. 
Besides, an adversary can guess the function of this layer according to the shape of the parameter, because different layers have different numbers of parameters (\eg two in the convolution layer) and different shapes of parameters (\eg $(3,3,64)$ in the convolution layer). 

To hide key model parameter information, we instead encapsulate parameters into their corresponding generated custom source codes of the obfuscated layer.
For example, for a simple one-layer feed-forward neural network, the output can be computed by $\mY = \sigma (\mW^T\mX + \vb ) $, where $\mX$, $\mW$, and $\vb$ is input tensor, parameter of the layer, and bias, respectively. 
$\sigma$ is the activation for neural nodes. 
For \toolname obfuscation, the network layer can be disguised as $\mY =  g(\mX) $, where $g$ is an unknown function. We then implement the correct computation (\ie $g$) in the generated custom TFLite source code, which we then obfuscate. 
%At the run,time, modified TFLite will identify the disguised layer and assign the correct parameter. 
At runtime, function $g$ will be invoked to achieve the computation from $\mX$ to $\mY$.
Now an adversary is unable to extract the key parameter information from our obfuscated model. 
For example, as shown in the {\color{green}\ding{193}} of Figure~\ref{fig:injection}, the explicit parameter information has been removed from the on-device DL model file.
Furthermore, the implementation of $g$ in the compiled DL library can be obfuscated using transitional and well-proven code obfuscation strategies~\citep{collberg2002watermarking}.
%we can use code obfuscation to hide the parameters in the modified source code.
Thus, adversaries will be hard to identify key model parameters by reverse engineering the compiled library.

\begin{figure*}[tb]
% \vspace{-0.5em}
  \begin{center}
    \includegraphics[width=0.85\textwidth]{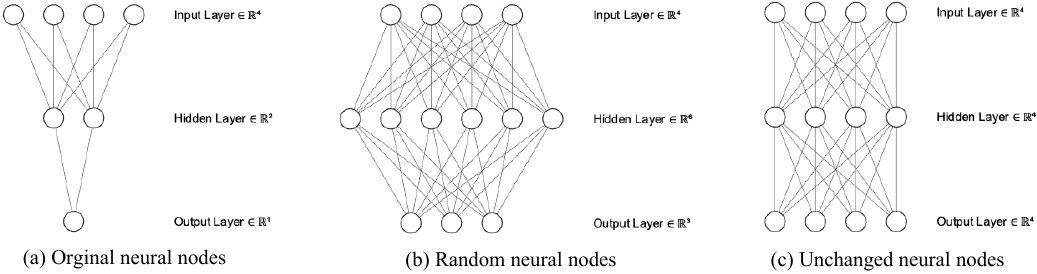}
  \end{center}
  \vspace{-1em}
  \caption{\ThirdObfuscateStrategyName for a simple feed-forward neural network.}
  \label{fig:neural_node}
  % \vspace{-0.5em}
\end{figure*}

% \vspace{-0.75em}
\paragraph{\textbf{\textit{\ThirdObfuscateStrategyName}}} Just obfuscating layer names and parameters is not enough, since an adversary may still infer the function of each layer according to the model structure.
For instance, Figure~\ref{fig:neural_node}(a) presents the structure of the neural network, where the input, hidden and output layers include four, two, and one node, respectively.
Attacker could search for a surrogate model according to the neural architecture.
To solve this problem, \toolname uses \textit{\thirdObfuscateStrategyName} to obfuscated neural architecture with the goal of confusing the adversary. 
We propose two strategies for network structure obfuscation: \textit{random} and \textit{align-to-largest}.
Given a model with output shapes $\vs = (\evs_0, \cdots, \evs_n)$, where $\evs_n$ refers to the number of dimensions for the $n$-th channel, the random strategy generates a random shape $\rvr = (\ervr_0, \cdots, \ervr_n)$ of the output for each layer.
Figure~\ref{fig:neural_node}(b) shows an obfuscated model of Figure~\ref{fig:neural_node}(a) using random strategy.
Second, the align-to-largest strategy finds the largest output shape $\vs'$ and then fills the output shapes of other layers to the size of $\vs'$.
Figure~\ref{fig:neural_node}(c) shows such an obfuscated model, where the output shapes of each layer are filled up to $(4)$.
Note that this will not affect the performance of models because the modified TFLite library will not compute the output using the provided neural structure information.

% \begin{minipage}[b]{0.64\linewidth}
% \vspace{-0.75em}
\paragraph{\textbf{\textit{Shortcut injection \& extra layer injection}}} 
%As we mentioned before, adversary can improve the effectiveness of black-box attacks by identifying the structure of target models.
Neural structure obfuscation changes the network structure by inserting new nodes, but the spatial relationships of original layers remain the same.
Therefore, even with the above three obfuscation strategies, the attacker can still infer node information by analyzing the spatial relationships of runtime data (\eg actual input-output values of each node).
To further obfuscate the model structure, hence, \toolname applies two more strategies: shortcut and extra layer injection. 
The injected shortcut and extra layers would destroy the original spatial relationships of TFLite models.

To automatically inject random shortcuts, \toolname first randomly selects a shortcut pair $(\rr_1, \rr_2)$.
The output index of $\rr_1$-th layer are then added to the input list of $\rr_2$-th layer.
% For example, the blue solid line in Figure~\ref{fig:injection}(b) is a shortcut inserted into the model extracted from a fruit recognition app, which connects the $2$-nd convolution layer and the $5$-th convolution layer.
For example, as shown in the {\color{blue}\ding{195}} of Figure~\ref{fig:injection} (b),  we add a shortcut between the layer `\texttt{Rwicza}' and `\texttt{Wjjzwh}' by adding the output index of `\texttt{Rwicza}' to the input list of `\texttt{Wjjzwh}'.
Second, to inject extra layers, just like the shortcut injection, \toolname randomly picks a layer pair $(\rr'_1, \rr'_2)$.
The input of the extra layer is the output node list of $\rr'_1$-th layer, and its output is added to the input list of $\rr'_2$-th layer.
For instance, an extra layer \texttt{Ypxumu} (as shown in the {\color{red}\ding{196}} of Figure~\ref{fig:injection}) is injected between the $3$-rd and $5$-th convolution layers.

Note that the injected shortcut and layers will not affect the prediction results of the deployed model because the modified TFLite library will ignore the  obfuscation part (\eg \texttt{Ypxumu} layers) in model inference.
Specifically, the extra layer $\mY = f(\mX)$ just needs to create the output with a specific shape to confuse the adversary.
In addition, extra layer injection will not significantly increase the latency of on-device models because the extra layer does not need many computational resources.

\paragraph{\textbf{\textit{Combing our five obfuscation strategies}}} 
To apply the proposed five on-device ML model obfuscation strategies sequentially, we designed an automatic model obfuscation tool \toolname. \toolname can automatically obfuscate an on-device ML model and produce compatible API libraries (\eg JAVA, Python) that can use it.
%The procedure of the model obfuscation by \toolname is shown in Procedure 1.
Figure~\ref{fig:injection} shows an example, where Figure~\ref{fig:injection}(a) is the original model from a real fruit recognition app, while Figure~\ref{fig:injection}(b) shows an example of the obfuscated model. We just add one shortcut and one extra layer to it. Note that developers can add more shortcuts and extra layers to achieve high obfuscation performance in practice.
The structure and parameters of the obfuscated model are quite different from that of the original model, which will make the obfuscated model hard to analyze.
To prevent an adversary from parsing the obfuscated model through reverse engineering the modified TFLite library and customized layers, we can use code obfuscation strategies, a well-developed technology to make the code unreadable and unable to be reverse engineered. Code obfuscation is a well-developed technology, so we do not show the detail in our paper. 
The combination of model and code obfuscation would hide the key information of models.

\subsection{Model Assembling \& Library Recompilation}
After obtaining the obfuscated model and modified TFLite source codes, \toolname assembles the new obfuscated model using the obfuscated model structure in Python. Then, it recompiles the modified TFLite library to support the newly generated obfuscated model. The newly generated obfuscated model and library can be packaged into mobile  apps or embedded device software to replace the original unobfuscated model and code library. To further prevent attackers from parsing the model by reverse engineering the modified TFLite library if it is possible, developers can use code obfuscation, which is a well-developed technique, to obfuscate the source code of the TFLite library.

% \vspace{-0.3em}
\section{\toolname Evaluation}

This work aims to use our proposed model obfuscation method to secure the deployed neural networks. To determine if this objective has been achieved, we evaluate \toolname by answering the following key research questions:

\begin{itemize}[leftmargin=*]

 \item \textbf{RQ1: } How effective is \toolname at obfuscating on-device ML models?

% \item \textbf{RQ2: } How effective is our approach in defending against the model extraction?

\item \textbf{RQ2: } What is the overhead of obfuscated on-device ML models?

\item \textbf{RQ3: }What is the impact of parameter sensitivity of \toolname  on obfuscated model overhead?

% \item \textbf{RQ4: } Which obfuscation strategy affects the size of produced DL library?

 \item \textbf{RQ4: } How effective is \toolname in defending against on-device ML model parsing?

% \item \textbf{RQ4: } How sensitive are the hyper-parameters?

\end{itemize}

% \vspace{-0.3em}

In this section, we show the evaluations of the proposed \toolname in terms of effectiveness, efficiency, and reliability. Our experiment settings are shown as follows:
% \vspace{-0.5em}
% \subsection{Experimental Setting}
% \vspace{-0.6em}
\paragraph{\textbf{Dataset}} To evaluate \toolname's performance on models with various structures for multiple tasks, we collected 10 TFLite models including a fruit recognition model, a skin cancer diagnosis model, MobileNet~\citep{howard2017mobilenets}, MNASNet~\citep{tan2019mnasnet}, SqueezeNet~\citep{iandola2016squeezenet}, EfficientNet~\citep{tan2019efficientnet}, MiDaS~\citep{ranftl2020towards}, LeNet~\cite{lecun1998gradient}, PoseNet~\citep{kendall2015posenet}, and SSD~\citep{liu2016ssd}. 
%We refer to them as model \ding{192} - \ding{201}, respectively. 
The fruit recognition and skin cancer diagnosis model were collected from Android apps (see the provided code repository). The other models were collected from the TensorFlow Hub~\footnote{\url{https://tfhub.dev/}}. All of these models can be found in our provided code repository.
% \vspace{-0.7em}
% \paragraph{Equipment} Although the on-device model runs on the edge devices, we evaluate the performance of proposed method on desktop because we just need to show the overhead of the \toolname. All experiments are run on the desktop with Intel(R) Xeon(R) W-2175 2.50GHz CPU. 
\paragraph{\textbf{Experimental Environment}} %The purpose of this \toolname evaluation is to demonstrate the differences in computational overhead between the original models and obfuscated models. 
%We evaluated the performance of 
\toolname is evaluated on a workstation with Intel(R) Xeon(R) W-2175 2.50GHz CPU, 32GB RAM, and Ubuntu 20.04.1 operating system.

\subsection{RQ1: Obfuscation Effectiveness}

\begin{table}[ht]
% \begin{table}[t]
\footnotesize
% \vspace{-1em}
    % \scriptsize
	\centering
 \caption{Comparison between original model files and obfuscated model files. `Operator codes': type of the neural layer in this model. `Tensors': the tensor that should be allocated in the memory. `Operators': each neural layer. `OR': obfuscation rate for original components. Note that the \textit{renaming} and \textit{parameter encapsulation} will increase and decrease the number of the corresponding model component, respectively.}
	\begin{tabular}{l|l|c|c}
 	\hline
             \multicolumn{4}{c}{\bf Operator codes} \\
             \hline
        &  \makecell[c]{Example} & Number & OR\\
	  % \multicolumn{2}{*}{\bf Operator codes} \\
		\hline
        Original
        &  "deprecated\_builtin\_code": 3 & 50 & \multirow{3}{*}{\bf 100\%} \\

        % & $\cdots$ {\color{red} 4 codes in total} \\
        \cline{1-3}
        \multirow{2}{*}{Obfuscated}
        & "deprecated\_builtin\_code": 32, & \multirow{2}{*}{510} & \\
        & "custom\_code": "Fvsfxf", & & \\
        % & "builtin\_code": "CUSTOM" & & \\

        % &$\cdots$ {\color{red} 9 codes in total} \\
        \hline\hline
             \multicolumn{4}{c}{\bf Tensors} \\
             % \hline
        % &  \makecell[c]{Example} & Count & Ratio\\
		\hline
        \multirow{5}{*}{Original}
        & "shape": [1,224,224,3], & \multirow{5}{*}{1345} & \multirow{7}{*}{\bf 100\%} \\
        & "name": "conv2d\_1/Relu", & &\\

        &$\cdots$ & & \\
        & "shape": [500,3200], & &\\
       &  "name": "dense\_1/kernel/transpose", & & \\
        % &$\cdots$ {\color{red} 16 tensors in total} \\
        \cline{1-3}
        \multirow{2}{*}{Obfuscated}
        & "shape": [1,224,224,3], & \multirow{2}{*}{534} & \\
        & "name": "Fvsfxf", &  & \\

        % &$\cdots$ {\color{red} 10 tensors in total}\\
        \hline\hline
             \multicolumn{4}{c}{\bf Operators} \\
	% \multicolumn{2}{c}{\bf Operators} & Count & Ratio \\
		\hline
        \multirow{4}{*}{Original}
        & "inputs": [3,2,0], & \multirow{4}{*}{510} & \multirow{8}{*}{\bf 100\%} \\
        & "outputs": [1], & & \\
        & "op\_type": "Conv2DOptions", & & \\
        & "builtin\_options": \{stride\_w: 1, $\cdots$\}, & & \\
        % &$\cdots$ {\color{red} 7 operators in total} \\
        \cline{1-3}
        \multirow{4}{*}{Obfuscated} 
        & "inputs": [0], & \multirow{4}{*}{510} & \\
        & "outputs": [1], & & \\
        & "op\_type": "Fvsfxf", & & \\
        & "custom\_options": \{84,0,1,3,1, $\cdots$\}, & & \\

        % &$\cdots$ {\color{red} 9 operators in total}\\
		\hline	
	\end{tabular}

 % \vspace{-2em}
 \label{tb:comparison}
% \end{table}
\end{table}

\begin{table*}[t]
\small
% \vspace{-1.0em}
\centering
\caption{Structure similarity between original and obfuscated structures. We use the Propagation Kernel algorithm to calculate the similarity between two graphs~\cite{JMLR:v21:18-370}. `$(n_1, n_2)$': obfuscated models with $n_1$ shortcuts and $n_2$ extra layers. The range of similarity is [0,1]. Note that we only use the \textit{extra shortcuts \& extra layer injection} in this experiment.}
\begin{tabular}{lcccccccccc|c}
\toprule
 $(n_1, n_2)$   & Fruit & Skin &MobileNet&MNASNet &SqueezeNet &EfficientNet &MiDaS &LeNet &PoseNet &SSD & \textbf{Average value} \\
\midrule
(10, 10)      & 0.78 & 0.81 & 0.80     & 0.95   & 0.89      & 0.93    &  0.97 &  0.60       & 0.81    & 0.95  & \textbf{0.86}        \\
% \hline
% (0, 0)  & 30.4 & 98.7 & 61.1     & 70.4   & 33.5      & 82.3        &  346.4  &   247.0        &  130.7   & 232.1  & \textbf{133.3}        \\
% (30, 0)& 30.7 & 97.9 & 62.3     & 72.7   & 33.1      & 82.0        &  346.9       &  247.1         &  130.2   &  231.0     &  \textbf{133.4}       \\
% % \hline
% (0, 10) & 30.2 & 98.6 & 61.1     & 74.1   & 34.9      & 82.3        &  346.7  &    251.5       & 130.9    & 231.6      & \textbf{134.2}        \\
% % \hline
% (0, 20) & 30.6 & 98.3 & 63.3     & 74.1   & 36.8      & 81.3        &  348.2  &    247.1       & 130.5    & 231.3      &  \textbf{134.2}       \\
% % \hline
(20, 20)  & 0.53 & 0.61 & 0.64     & 0.82   & 0.75      & 0.84    &  0.95 &  0.59       & 0.65    &   0.89    & \textbf{0.74}        \\
(30, 30)  & 0.48 & 0.59 & 0.51     & 0.76   & 0.63      & 0.77    &  0.92 &  0.59       & 0.64    &  0.81     & \textbf{0.67}        \\
% (0, 20)  & - & - & -     & -   & -      & -    &  - &  -       & -    & -  & \textbf{-}        \\
\bottomrule
\end{tabular}
\label{tb:stru_sim}
\end{table*}

\begin{table}[t]
\footnotesize
% \vspace{-1.0em}
\centering
\caption{Structure similarity between the original models. `\ding{192} - \ding{201}': Fruit, Skin, MobileNet,MNASNet, SqueezeNet, EfficientNet, MiDaS, LeNet, PoseNet, SSD. }
\begin{tabular}{l|cccccccccc}
\toprule
 \multicolumn{1}{c}{}  & \ding{192} & \ding{193} &\ding{194}&\ding{195} &\ding{196} &\ding{197} &\ding{198} &\ding{199} &\ding{200} &\ding{201}  \\
\hline
\ding{192}  & 1.0 & 0.99 & 0.98     & 0.80   & 0.66      & 0.80    &  0.66 &  0.18       & 0.97    & 0.67         \\
\ding{193}  & 0.99 & 1.0 & 0.99     & 0.81   & 0.68      & 0.82    &  0.68 &  0.18       & 0.97    & 0.69            \\
\ding{194}  & 0.98 & 0.99 & 1.0     & 0.82   & 0.67      & 0.83    &  0.68 &  0.18       & 0.97    & 0.68             \\
\ding{195}  & 0.80 & 0.81 & 0.82    & 1.0    & 0.63      & 0.99    &  0.94 &  0.18       & 0.79    & 0.58         \\
\ding{196}  & 0.66 & 0.68 & 0.67    & 0.63   & 1.0       & 0.69    &  0.57 &  0.20       & 0.69    & 0.63            \\
\ding{197}  & 0.80 & 0.82 & 0.83    & 0.99   & 0.68      & 1.0     &  0.93 &  0.18       & 0.81    & 0.66             \\
\ding{198}  & 0.66 & 0.68 & 0.68    & 0.94   & 0.57      & 0.93    &  1.0  &  0.12       & 0.68    & 0.50         \\
\ding{199}  & 0.18 & 0.18 & 0.18    & 0.18   & 0.20      & 0.18    &  0.12 &  1.0        & 0.55    & 0.62            \\
\ding{200}  & 0.97 & 0.97 & 0.97    & 0.79   & 0.69      & 0.81    &  0.68 &  0.55       & 1.0    & 0.83             \\
\ding{201}  & 0.67 & 0.69 & 0.68    & 0.58   & 0.63      & 0.66    &  0.50 &  0.62       & 0.83    & 1.0         \\

% (0, 20)  & - & - & -     & -   & -      & -    &  - &  -       & -    & -  & \textbf{-}        \\
\bottomrule
\end{tabular}
\label{tb:stru_sim_ori}
\end{table}

In addressing this RQ we aim to demonstrate the effectiveness of \toolname. 
Attackers can either extract ML model information from the model file~\cite{li2021deeppayload} or analyze the structure of models to collect similar models from the web for security exploitation~\cite{huang2021robustness,huang2022smart}. Thus, we demonstrate the effectiveness of \toolname in obfuscating model files and structures.
% and showcase the differences between the original on-device ML model and the obfuscated ML model, respectively. 
% Note that the \textit{renaming}, \textit{parameter encapsulation}, and \textit{\thirdObfuscateStrategyName} approaches of \toolname are employed in model file obfuscation, while the other two strategies, \textit{shortcut injection} and \textit{extra layer injection}, are used to obfuscate the model structure. 
% We use LeNet~\citep{lecun1998gradient} in this case study. It is a 7-layer neural network and has four different neural layers, including \texttt{Conv2D}, \texttt{Maxpool2D}, \texttt{FullyConnected}, and \texttt{Logistic}.

\paragraph{\textbf{Model File Obfuscation}} In this experiment, \textit{renaming}, \textit{parameter encapsulation}, and \textit{\thirdObfuscateStrategyName} approaches of \toolname are employed. To demonstrate the effectiveness of these approaches, we use \textit{Flatc}\footnote{\url{https://google.github.io/flatbuffers/flatbuffers_guide_use_python.html}}, an existing python tool to parse the information of TFLite models in the buffer.
% First, attackers may extract ML model information from the model file. Hence, we use the existing python tool, \textit{Flatc}\footnote{\url{https://google.github.io/flatbuffers/flatbuffers_guide_use_python.html}}, to parse the information of models in the buffer.

A comparison of file contents between the original models and \toolname obfuscated models is shown in Table \ref{tb:comparison}. The results include the summary of all 10 models.
% Note that we use the \textit{renaming}, \textit{parameter encapsulation}, and \textit{\thirdObfuscateStrategyName} to show how effective our \toolname for obfuscating the model file data in this experiment.
For operator codes (types of the neural layer), \toolname creates different obfuscated operator names (e.g., layer names) for each neural layer. 
As shown in the first section of Table \ref{tb:comparison}, the original model assigns the same name to the same type of layers (\ie 510 layers are assigned with 50 names), while the obfuscated models assign random names to each of the individual layer (i.e., 510 different operator codes with random names).
This can confuse attackers in inferring the functionality of layers by hiding the relationship between certain layers. %by how often the layers are used.

For tensors, as shown in the second section of Table \ref{tb:comparison}, the original model file contains the exact shape and name of each allocated tensor. The allocated tensors include the inputs, outputs, and weights of each layer. It also provides the index of layers' weights. 
In contrast, the obfuscated model has a fixed tensor shape ( \texttt{[1,224,224,3]} in the example) and random names. In addition, it does not have information on layers' weights. So, the obfuscated model file only has 534 allocated tensors. Attackers cannot extract the weights and output shape of each layer from model files.
% \textcolor{red}{(one sentence here to explain why/how obfuscated tensors are more resilience to attacks.)}

For operators, as shown in the third section of Table \ref{tb:comparison}, the original model file contains detailed information about each layer, including the inputs, outputs, type, and settings (`\texttt{builtin\_options}'). In contrast, the obfuscated models exclude the weights (\texttt{inputs 3 and 2} in this example) of layers in the input list, hide the operator types, and randomize the settings. Thus, attackers cannot get the details of layers. Overall, these three obfuscation strategies (\ie \textit{renaming}, \textit{parameter encapsulation}, and \textit{\thirdObfuscateStrategyName}) can effectively obfuscate the model file. \textbf{\toolname can confuse attackers, especially those using automatic tools or reverse engineering~\cite{li2021deeppayload}, to extract model information.}

\begin{figure}[!tb]
% \vspace{-0.5em}
  \begin{center}
    \includegraphics[width=0.45\textwidth]{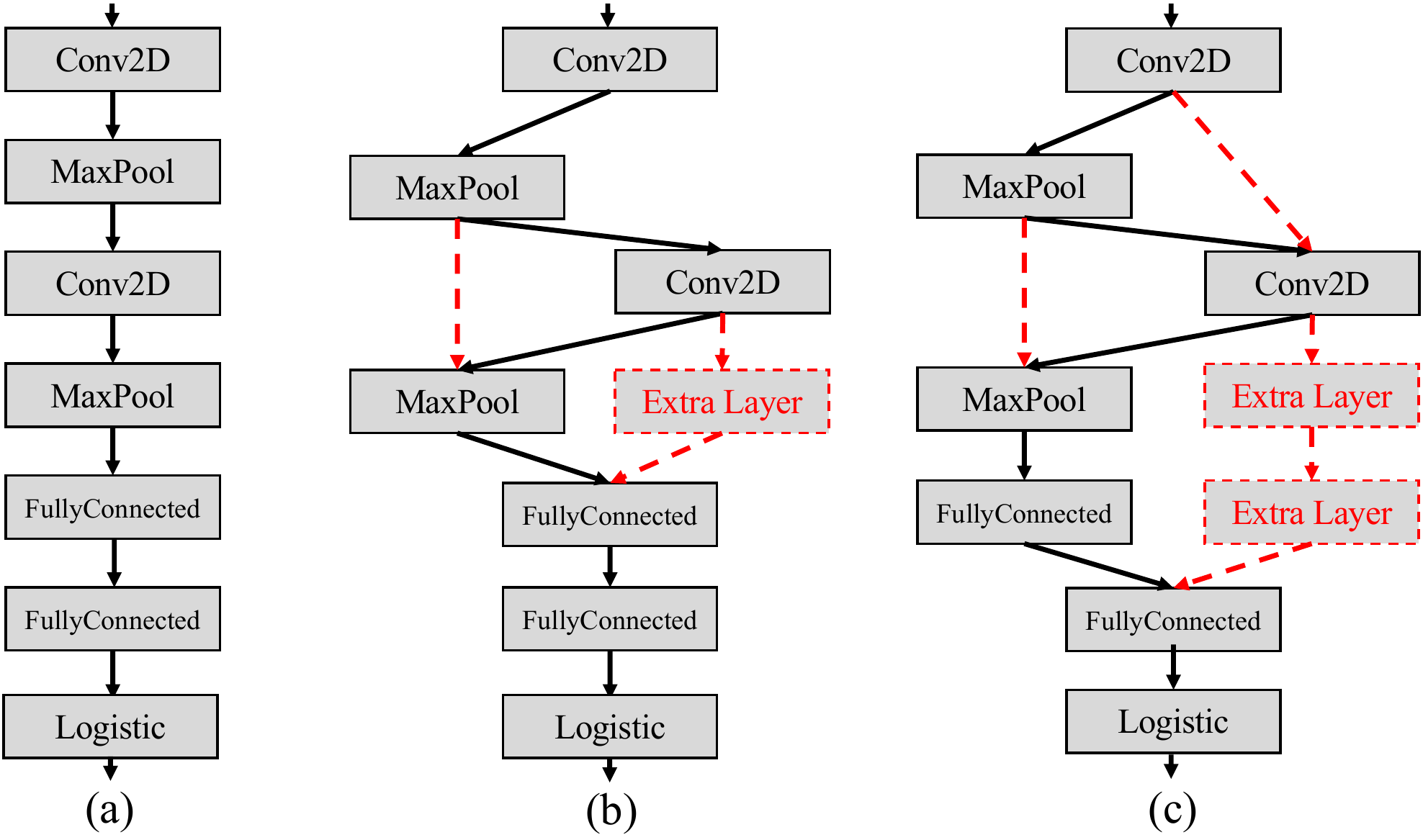}
  \end{center}
  % \vspace{-0.5em}
  \caption{Example of structure obfuscation for LeNet. (a) Original model. (b) Obfuscated model with one extra shortcut \& layer. (c) Obfuscated model with two extra shortcuts \& layers. Red dotted line and red dotted block represent extra shortcuts and extra layers, respectively.  }
  \label{fig:structure_obf}
  % \vspace{-0.5em}
\end{figure}

\paragraph{\textbf{Model Structure Obfuscation}} While we can obfuscate the model file component, attackers still can parse the model using model structure. In order to mitigate against such attacks, we conducted a second experiment to showcase the effectiveness of \toolname in obfuscating model structures, which only utilizes the \textit{shortcut} and \textit{extra layer injection}.
% Second, attackers can analyze the features of models to collect similar models from the web for security exploitation~\cite{huang2021robustness,huang2022smart}. They do not need precise model information to reconstruct the original model. Although we can obfuscate the model file component, attackers still can parse the model using model structure. To demonstrate the effectiveness of \toolname in obfuscating the model structure, we only apply the \textit{shortcut} and \textit{extra layer injection} in this experiment. 
We use the Propagation Graph Kernel method~\cite{JMLR:v21:18-370} to calculate the structure similarity between graphs (\ie original model structure and obfuscated model structure), which is shown in Table~\ref{tb:stru_sim}. 
% we show a visualization of LeNet before and after applying \textit{Shortcut injection \& extra layer injection}, which is shown in Figure~\ref{fig:structure_obf}. We set the number of extra shortcuts and extra layers as two. 
We utilize the structure similarity between the models to demonstrate the effectiveness of our approach in obfuscating the model structure. This is because the model structure can be represented as a directed graph, as shown in Figure~\ref{fig:structure_obf}, with each layer functioning as a node in the graph.
% Then, to evaluate the effectiveness of our method in obfuscating the model structure, we only apply the \textit{shortcut} and \textit{extra layer injection} in this experiment. We use the Propagation Graph Kernel method to calculate the structure similarity between graphs (\ie original model structure and obfuscated model structure)~\cite{JMLR:v21:18-370}, which is shown in Table~\ref{tb:stru_sim}.
% As can be seen, \textbf{it is difficult to recognize the shortcut or extra layer without prior knowledge of the model structure}. When a large number (\eg more than 30) of shortcuts and extra layers is used to obfuscate the TFLite, the structure of the obfuscated model will become extremely confused. 

Table~\ref{tb:stru_sim} and \ref{tb:stru_sim_ori} show the structure difference between the original model and the obfuscated model will lay on the normal range of the difference (\ie 0.50-0.95) between original models when the number of extra shortcuts \& extra layers is 20.
\textbf{Our results show  that attackers will find it hard to use structure similarity to identify the correct similar models (\ie a model that has similar structure and parameters) from the web when the model information -- structure and parameters -- has been obfuscated by \toolname.} In addition, the more the number of extra shortcuts and extra layers, the higher the effectiveness of structure obfuscation. When the model has higher complexity~(\eg MiDaS), \toolname needs more number of extra shortcuts \& layers to generate obfuscated models that have low structure similarity with original models.

\begin{tcolorbox}[colback=gray!5!white,colframe=gray!85]
Overall, the \toolname can effectively obfuscate both the data and structures of on-device ML models. 
\end{tcolorbox}

% It is hard for automatic tools for analyzing the model and human experts to extract the information from the obfuscated model.

% In experiments, we will answer the following three research questions: (1) How is the effectiveness of the proposed model obfuscation? (2) How is the overhead of \toolname? (3) How is the compatibility of the \toolname for different TensorFlow and TFLite version?

% \begin{table*}[t]
% % \small
% \centering
% % \vspace{-1em}
% \caption{The obfuscation error of the proposed model obfuscation method. `Error': the output difference between the original model and the obfuscated model.}
% \begin{tabular}{lccccccccccc}
% \toprule
%                &\ding{192}&\ding{193}&\ding{194}&\ding{195} &\ding{196} &\ding{197} &\ding{198} &\ding{199} &\ding{200} &\ding{201} & \textbf{Average} \\
% \midrule
% Error       & 0.0 & 0.0 & 0.0     & 0.0   & 0.0      & 0.0    &  0.0 &  0.0       & 0.0    & 0.0  &  0.0       \\
% \bottomrule
% \end{tabular}
% % \vspace{-1em}
% \label{tb:error}
% \end{table*}

\begin{table*}[t]
\small
% \vspace{-1.0em}
\centering
\caption{Time overhead (seconds per 1000 inputs) of the original model and obfuscated model. We use five obfuscation strategies. We set the number of extra shortcuts and extra laters to 20. `+' and `-' refer to the increase and decrease, respectively.}
\begin{tabular}{lcccccccccc|c}
\toprule
               & Fruit & Skin &MobileNet&MNASNet &SqueezeNet &EffcientNet &MiDaS &LeNet &PoseNet &SSD & \textbf{Average value} \\
\midrule
Original      & 30.3 & 98.5 & 62.1     & 72.5   & 33.4      & 81.2    &  346.6 &  1.5       & 130.4    & 231.2  & \textbf{108.8}        \\
% \hline
% (0, 0)  & 30.4 & 98.7 & 61.1     & 70.4   & 33.5      & 82.3        &  346.4  &   247.0        &  130.7   & 232.1  & \textbf{133.3}        \\
% (30, 0)& 30.7 & 97.9 & 62.3     & 72.7   & 33.1      & 82.0        &  346.9       &  247.1         &  130.2   &  231.0     &  \textbf{133.4}       \\
% % \hline
% (0, 10) & 30.2 & 98.6 & 61.1     & 74.1   & 34.9      & 82.3        &  346.7  &    251.5       & 130.9    & 231.6      & \textbf{134.2}        \\
% % \hline
% (0, 20) & 30.6 & 98.3 & 63.3     & 74.1   & 36.8      & 81.3        &  348.2  &    247.1       & 130.5    & 231.3      &  \textbf{134.2}       \\
% % \hline
Obfuscated & 30.2 & 98.7 & 63.2     & 74.3   & 34.6      & 82.9        &  349.2  &    1.6        & 130.4    & 237.1      &  \textbf{110.2}       \\
\hline
Difference & -0.1 &  +0.2 & +1.1     & +1.8   & +1.2      & +1.7        &  +2.6  &    +0.1        & 0.0    & +5.9      &  \textbf{+1.4}       \\
\bottomrule
\end{tabular}
% \vspace{-0.5em}
\label{tb:latency}
\end{table*}

\begin{table*}[t]
 \small
\centering
\caption{Overhead of the model obfuscation on random access memory (RAM) cost (Mb per model). We use five obfuscation strategies. To eliminate the influence of other processes on the test machine, we show the increment of RAM usage. }
\begin{tabular}{lcccccccccc|c}
\toprule
               & Fruit & Skin &MobileNet&MNASNet &SqueezeNet &EffcientNet &MiDaS &LeNet &PoseNet &SSD & \textbf{Average value} \\
\midrule
Original       & 18.8 & 49.5 & 33.1     & 46.5   & 41.3      & 51.9        &   237.1      &  2.8         &  43.4   &  97.8     &  \textbf{62.2}       \\
% % \hline
% (0, 0)  & 18.9 & 53.3 & 33.1     & 49.9   & 45.5      & 56.8        &   244.8      &    381.2       &  40.6   &  99.8     &   \textbf{102.4}      \\
% (30, 0)& 18.7 & 51.4 & 32.6    & 49.5   & 46.0      & 55.8        &   244.6       &    381.6       &  40.3   &    100.2   &   \textbf{102.0}      \\
% % \hline
% (0, 10) & 25.9 & 58.8 & 42.5     & 53.6   & 49.2      & 69.5        &   248.1      &   386.7         &  55.0   &  107.5     &  \textbf{109.7}       \\
% % \hline
% (0, 20) & 30.0 & 67.3 & 50.4     & 56.7   & 56.0      & 66.2        &   253.1      &   387.7        &  66.9   & 112.0      &  \textbf{114.6}       \\
% % \hline
Obfuscated & 30.1 & 65.6 & 50.5     & 56.6   & 52.6      & 66.3        &   254.5      &   3.1        & 61.6    &  107.6     &  \textbf{74.8}       \\
\hline
Difference & +11.3 & +16.1 & +17.4     & +10.1   & +11.3      &   +14.4      &   +17.4      &  +0.3         & +18.2    &  +9.8     &  \textbf{+12.6}       \\

\bottomrule
\end{tabular}
% \vspace{-0.5em}
\label{tb:mem}
\end{table*}

\begin{table*}[t]
\small
% \vspace{-0.5em}
\centering
\caption{Size change (Mb) of the TFLite library and models after the obfuscation.
The size of the obfuscated model is reduced to a few Kb.
%The reduced size of TFLite models is the original size because obfuscated models just have a few Kb data.
The original library size is 183 Mb.}
\begin{tabular}{lcccccccccc}
\toprule
               & Fruit & Skin &MobileNet&MNASNet &SqueezeNet &EffcientNet &MiDaS &LeNet &PoseNet &SSD \\
\midrule
Library (renaming)       & +8.7 & +31.9 & +19.6     & +38.7   & +8.8      & +40.6       &   +126.1   &  +11.6     &  +9.7 & +52.1     \\
Library (all strategies)       & +8.7 & +31.9 & +19.6     & +38.7   & +8.8      & +40.6        &   +126.1   &  +11.6     &  +9.7   & +52.1            \\
Model         & -5.5 & -16.9 & -10.3     & -17.5   & -5.0      & -18.6        &   -66.3      &  -6.5         &  -5.0   & -27.5             \\
\midrule
  Total        & +3.2  & +15.0  & +9.3   & +21.2   & +3.8      & +22.0        &  +59.8      &  +5.1         &  +4.7   &  +24.4    \\
\bottomrule
\end{tabular}
\label{tb:pythonsize}
\end{table*}

\subsection{RQ2: Obfuscation Overhead}
% In this subsection, we evaluate the efficiency of \toolname obfuscated models. 
Ideally, applying \toolname to an on-device ML model should not affect the prediction accuracy of the original models, while still providing sufficient defense against model attacks. To this end, we apply all five proposed obfuscation strategies to each model and compare the prediction results based on 1,000 randomly generated inputs. The obfuscation error is calculated as $ ||\vy - \vy'||_2 $, where $\vy$ and $\vy'$ is the output of original models and obfuscated models, respectively. Note that the number of extra layers and shortcuts is set to 30 in \textit{shortcut injection \& extra layer injection}. In this experiment, \textbf{the obfuscation error of the proposed model obfuscation method is 0 for all 10 on-device models. It shows that our obfuscation strategies have no impact on the prediction results of the original models}.

In order to evaluate the efficiency impact of our obfuscation strategies on ML models, we conducted experiments to measure the runtime overhead of each model in our dataset. We measured both the time and memory overhead of the proposed obfuscation method under various settings, based on 1,000 randomly generated instances. The results of these experiments are presented in Tables \ref{tb:latency} and \ref{tb:mem}, which respectively report the time overhead and memory overhead of the obfuscated models.
% `$(n_1, n_2)$' in the table indicates the the number of shortcuts ($n_1$) and number of extra layers ($n_2$) applied, where $(0, 0)$ indicates only the basic obfuscations (\ie renaming, parameter encapsulation) and \thirdObfuscateStrategyName are involved. 
We also include the time and memory consumption of the original models as the baseline. As shown in Table~\ref{tb:latency}, even though extra layers are injected into the obfuscated model, \textbf{\toolname obfuscated models incur a negligible time overhead} (\ie approximately 1\% on average for the most time-consuming obfuscation). The differences between using various obfuscation settings are also not significant. Because the \textit{parameter encapsulation} will remove some data processing steps in the source code of APIs, the basic obfuscation (`(0,0)' in Table~\ref{tb:latency}) may reduce the latency of TFLite models.

% To evaluate the overhead of the proposed model obfuscation, we test the latency of the original TFLite model and obfuscated TFLite model, which is shown in Table~\ref{tb:latency}. The experiment shows all obfuscation strategies will not significantly increase the latency of the on-device model. It is because the extra layer create the output with a specific shape.

The memory overhead for \toolname obfuscated models is shown in Table~\ref{tb:mem}. To eliminate the impact of different memory optimization methods, we use peak RAM usage where the model preserves all intermediate tensors. 
% It can be seen that the other obfuscation strategies do not affect memory usage except for the \textit{extra layer injection}. 
% Memory usage increases when the number of extra layers increases. Therefore, a trade-off between the complexity of the obfuscation and the memory overhead is worth considering when choosing the obfuscation strategy. 
We argue that even with 20 extra shortcuts and 20 extra layers in our experiment, which provides sufficient protection to original models, \textbf{the memory overhead of obfuscated ML models is acceptable (approximately 20\%)}.

Considering that the models are deployed on mobile devices that have limited storage space, we also present the size differences of the modified TFLite software library and the obfuscated models. Note that the size change is caused by creating additional \texttt{.so} files to support the inference of the obfuscated layer. Hence, the size difference will be similar with different implementations (\eg Python, Java, etc.). Table~\ref{tb:pythonsize} shows the size change to the TFLite Python library and the TFLite models after applying \toolname obfuscation strategies. We use all obfuscation strategies and the number of extra shortcuts and extra layers is 30.
Our results show that \textbf{the library size change is mainly caused by the \textit{renaming} method}, because it will create a new API for the renamed layer in the TFLite library. In addition, the \textit{extra layer injection} will not increase the size of the library, although it will create a new API to support the extra layer. Because the extra layer just has a simple function, the effect of the \textit{extra layer injection} in size is negligible. 
Our obfuscation strategies \textbf{significantly reduce the size of the TFLite model file} because it only keeps the obfuscated minimal structure information. However, \textbf{the size of the TFLite library is significantly increased}, and increases the size of the application deployed on the mobile device. 

\begin{tcolorbox}[colback=gray!5!white,colframe=gray!85]
The time overhead of \toolname on on-device ML models is negligible. In addition, the memory overhead on obfusacted models is acceptable  (approximately 20\%) when the model has sufficient protection. However, \toolname will increase the size of the TFLite library.  
\end{tcolorbox}

\begin{table*}[t]
 \small
% \vspace{-1.0em}
\centering
\caption{Time consumption (seconds per 1000 inputs) of obfuscated models in different settings. `$(n_1, n_2)$': obfuscated models with $n_1$ shortcuts and $n_2$ extra layers.}
% \vspace{-0.5em}
\begin{tabular}{lccccccccccc}
\toprule
 ($n1$, $n2$)              & Fruit & Skin &MobileNet&MNASNet &SqueezeNet &EffcientNet &MiDaS &LeNet &PoseNet &SSD & \textbf{Average} \\
\midrule
% \hline
(0, 0)  & 30.4 & 98.7 & 61.1     & 70.4   & 33.5      & 82.3        &  346.4  &   1.5        &  130.7   & 232.1  & \textbf{108.7}        \\
(10, 0)  & 30.4 & 98.7 & 61.0     & 70.7   & 33.4      & 82.1        &  346.5  &   1.7        &  130.2   & 232.1  & \textbf{108.7}        \\
(20, 0)  & 30.5 & 98.6 & 61.4     & 70.4   & 33.6      & 82.4        &  346.4  &   1.6        &  130.6   & 232.3  & \textbf{108.8}        \\
(30, 0)& 30.7 & 97.9 & 62.3     & 71.7   & 33.1      & 82.0        &  346.9       &  1.5         &  130.2   &  231.0     &  \textbf{108.7}       \\
% \hline
(0, 10) & 30.2 & 98.6 & 61.1     & 74.1   & 34.9      & 82.3        &  346.7  &   1.6      & 130.9    & 231.6      & \textbf{109.2}        \\
% \hline
(0, 20) & 30.6 & 98.3 & 63.3     & 74.1   & 36.8      & 81.3        &  348.2  &    1.6       & 130.5    & 231.3      &  \textbf{109.6}       \\
% \hline
(0, 30) & 30.4 & 98.5 & 63.2     & 74.6   & 34.5      & 82.8        &  349.0  &    1.8        & 130.4    & 236.7      &  \textbf{110.2}       \\
\bottomrule
\end{tabular}
% \vspace{-0.5em}
\label{tb:latency_para}
\end{table*}

\begin{table*}[t]
 \small
\centering
\caption{Random access memory (RAM) cost (Mb per model) of obfuscated models in different settings. `$(n_1, n_2)$': obfuscated models with $n_1$ shortcuts and $n_2$ extra layers.}
% \vspace{-0.5em}
\begin{tabular}{lccccccccccc}
\toprule
 ($n1$, $n2$)              & Fruit & Skin &MobileNet&MNASNet &SqueezeNet &EffcientNet &MiDaS &LeNet &PoseNet &SSD & \textbf{Average} \\
\midrule
% Original       & 18.8 & 49.5 & 33.1     & 46.5   & 41.3      & 51.9        &   241.1      &  373.8         &  39.4   &  97.8     &  \textbf{99.3}       \\
% \hline
(0, 0)  & 18.9 & 53.3 & 33.1     & 49.4   & 45.5      & 56.4        &   244.8      &    2.9       &  40.6   &  99.9     &   \textbf{64.5}      \\
(10, 0)  & 18.8 & 53.2 & 33.3     & 49.7   & 45.8      & 56.8        &   244.8      &   2.8       &  40.5   &  99.8     &   \textbf{64.6}      \\
(20, 0)  & 18.9 & 53.5 & 32.8     & 49.7   & 45.9      & 56.7        &   244.3      &   2.8       &  40.5   &  99.9     &   \textbf{64.5}      \\
(30, 0)& 18.7 & 53.4 & 32.6    & 49.5   & 46.0      & 55.8        &   244.6       &    2.9       &  40.3   &    100.2   &   \textbf{64.4}      \\
% \hline
(0, 10) & 25.9 & 58.8 & 42.5     & 53.6   & 49.2      & 69.5        &   248.1      &   2.9         &  55.0   &  107.5     &  \textbf{71.3}       \\
% \hline
(0, 20) & 30.0 & 67.3 & 50.4     & 56.7   & 52.0      & 66.2        &   253.1      &   3.1        &  63.9   & 110.0      &  \textbf{75.2}       \\
% \hline
(0, 30) & 35.9 & 81.9 & 55.2     & 57.3   & 71.6      & 73.0        &   267.7      &   3.1        & 73.4    &  123.8     &  \textbf{81.9}       \\

\bottomrule
\end{tabular}
% \vspace{-0.5em}
\label{tb:mem_para}
\end{table*}

\begin{table*}[t]
% \vspace{-1.0em}
 \small
\centering
%\caption{The success number of existing model conversion tools and attack extracting the information of models. `\textit{Basic obfuscation}': \textit{renaming} + \textit{parameter encapsulation}.}
\caption{The success number of existing reverse engineering methods to extract the information of on-device models with different obfuscation strategies. `$\surd$': this model parsing method cannot extract information for all models.} 
% \vspace{-0.5em}
\begin{tabular}{l|ccc|c|c}
\toprule
                                       &  \multicolumn{3}{c|}{\bf Model conversion tool} & \multicolumn{1}{c|}{\bf Parsing model in the buffer}  & {\bf Feature analyzing} \\
\hline
                                       & TF-ONNX & TFLite2ONNX & TFLite2TF & Reverse Engineering in FlatBuffer~\cite{li2021deeppayload} & Smart App Attack~\cite{huang2022smart} \\
\hline
Without obfuscation                      & 10              & 9          & 9                 & 10           & 8         \\ 
% \hline
\textit{Renaming}                   & $\surd$               & $\surd$           & $\surd$                 & $\surd$            & 8          \\
\textit{Parameter encapsulation}     & $\surd$               & $\surd$           & $\surd$                 & $\surd$            & 2          \\ 
\textit{\ThirdObfuscateStrategyName}    & $\surd$               & $\surd$           & $\surd$                 & $\surd$            & 8          \\ 
\textit{Shortcut injection}         & $\surd$               & $\surd$           & $\surd$                 & $\surd$            & 8          \\ 
\textit{Extra layer injection}     & $\surd$               & $\surd$           & $\surd$                 & $\surd$            & 8          \\ 
\hline
\textbf{\textit{\toolname}}     & $\surd$               & $\surd$           & $\surd$                 & $\surd$            & $\surd$          \\ 
% \midrule 
\hline
\end{tabular}
% \vspace{-1em}
\label{tb:extraction}
\end{table*}
% \subsection{Q4: Utility of \toolname}

\subsection{RQ3: Parameter Sensitivity}
In this study, our research question investigates the overhead of obfuscated on-device models in various settings. Our proposed tool, \toolname, includes two hyper-parameters: the number of additional shortcuts and the number of extra layers, which are used to regulate the effectiveness of obfuscating model structures. As demonstrated in Tables \ref{tb:latency_para} and \ref{tb:mem_para}, we conduct ablation studies to examine the impact of these hyper-parameters on the overhead of obfuscated models.
The notation $(n_1, n_2)$ in the table represents the number of shortcuts ($n_1$) and extra layers ($n_2$) applied. When $(0, 0)$ is used, it indicates that only basic obfuscations (\ie \textit{renaming}, \textit{parameter encapsulation}) and \textit{\thirdObfuscateStrategyName} are employed. Our results in Table~\ref{tb:latency_para} indicate that \textbf{these two parameters have little effect on the time overhead}. However, as shown in Table~\ref{tb:mem_para}, \textbf{the RAM cost of obfuscated models increases with the number of extra layers}.
% \textbf{Therefore, a trade-off between the complexity of the obfuscation and the memory overhead is worth considering when choosing the obfuscation strategy}. 

\begin{tcolorbox}[colback=gray!5!white,colframe=gray!85]
% \toolname paramters impact obfuscated model memory performance and should be considered when using the tool.
Hence, it is worthwhile to consider the trade-off between the obfuscation complexity and the memory overhead.
\end{tcolorbox}

\subsection{RQ4. Resilience to Attacks}
\label{RQ4}
To demonstrate the effectiveness in concealing information of models, we use three distinct model parsing methods: 

\begin{enumerate}[leftmargin=*]
    \item The first method involves model format conversion. Existing model conversion tools utilize public APIs to access a model's structure and weights, and then assemble them into a new model with different formats. We utilize three tools, namely TF-ONNX~\citep{tf2onnx}, TFLite2ONNX~\citep{tflite2onnx}, and TFLite2TF~\citep{tflite2tensorflow}, to evaluate the performance of \toolname. If a tool can convert the model format, we consider it a successful model information extraction. Conversely, if \toolname is effective, these tools will be unable to extract the model information.
    \item The second method involves reverse engineering in the buffer. The on-device model format TFLite loads data on buffer~\footnote{\url{https://google.github.io/flatbuffers/}}. We refer to a study by Li et al.~\citep{li2021deeppayload}, in which the researchers attempt to extract a model's structure and weights in FlatBuffer. If \toolname is effective, the FlatBuffer extractor will be unable to parse the data of the obfuscated model and reverse it to the original one.
    \item The third method involves finding a similar differentiable model from the Internet by comparing the features among models. Although the mainstream on-device model format TFLite does not support some advanced functions like auto-differentiation to protect the deployed model, attackers such as the App Attack can find a differentiable model with a similar structure and weights from the Internet~\citep{huang2022smart}. For the App Attacking method, if it can correctly identify the obfuscated model that has the same model structure as the original one on TensorFlow Hub, we consider it a successful extraction of the model information.
\end{enumerate}

In this experiment, we evaluated the effectiveness of the proposed obfuscation strategies, including \textit{renaming}, \textit{parameter encapsulation}, \textit{\thirdObfuscateStrategyName}, and \textit{shortcut \& extra layer injection}, on the original models.
Table~\ref{tb:extraction} shows the results of our evaluation, where we used three different model parsing methods to attempt to extract information from the TFLite models. \textbf{Our proposed \toolname successfully prevented all existing model parsing tools and methods from obtaining the information of TFLite models.}
In addition, we found that parameter encapsulation alone was able to prevent the App Attack from finding surrogate models on six of the models. However, applying other obfuscation strategies separately did not confuse the App Attack, as it uses the features of the model structure and parameters simultaneously to analyze the model. Therefore, \textbf{to defend against the App Attack, we need to obfuscate the model structure and parameters simultaneously by combining the proposed obfuscation methods.}

\begin{tcolorbox}[colback=gray!5!white,colframe=gray!85]
The proposed \toolname is effective in defending against all three types of model parsing methods, thus providing robust protection for deployed models.
\end{tcolorbox}

\section{Discussion}
In this section, we will present the taxonomy of model obfuscation strategies and the limitation of our methods. In addition, we show the possible way to extract information from the obfuscated model.
% \vspace{-0.7em}
\begin{figure}[t]
% \vspace{-1em}
  \begin{center}
    \includegraphics[width=0.99\linewidth]{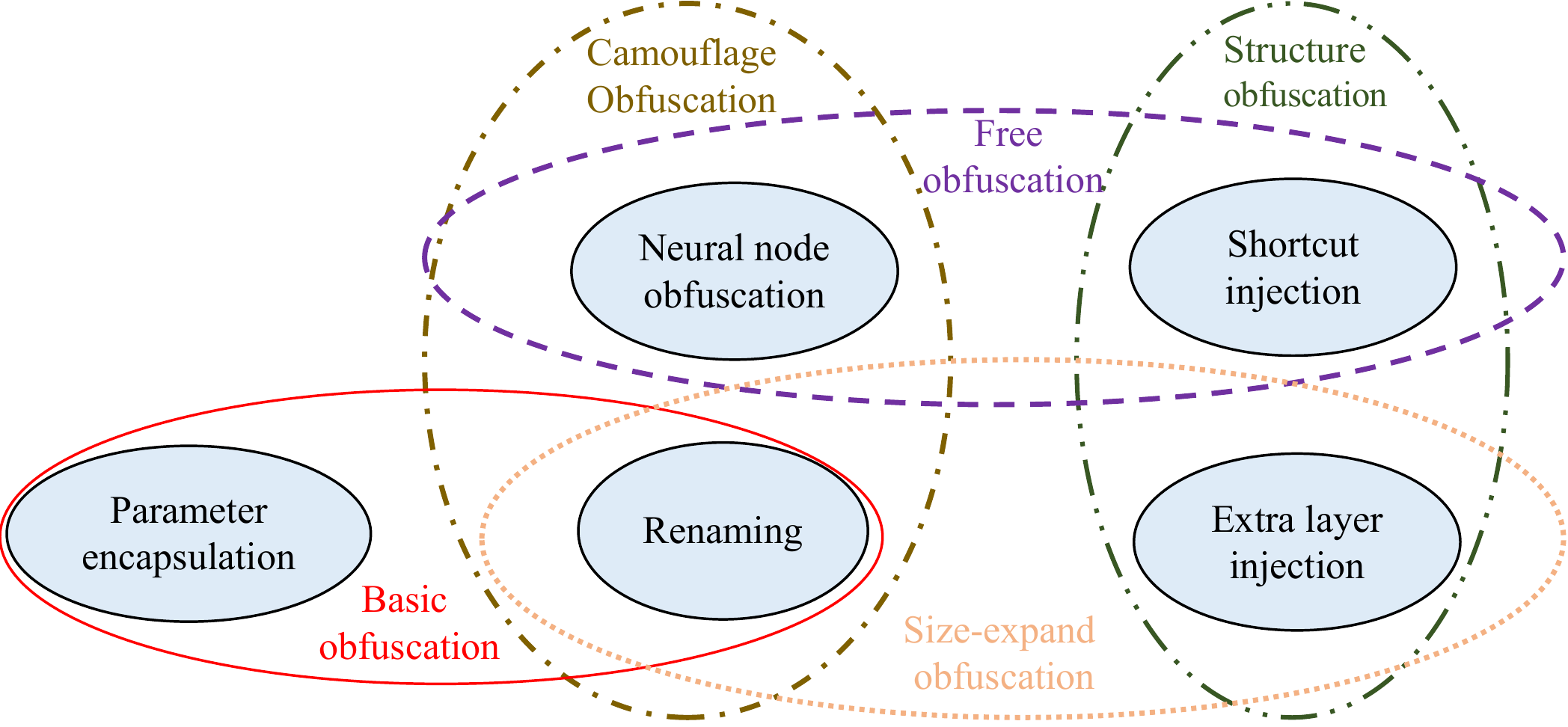}
  \end{center}
  % \vspace{+1em}
  % \vspace{-0.7em}
  \caption{Taxonomy of model obfuscation strategies.}
  \label{fig:taxonomy}
  % \vspace{-0.8em}
  
\end{figure}
% \vspace{-1em}
\subsection{Taxonomy of Model Obfuscation Strategies}
In this paper, we proposed five different model obfuscation strategies. Figure~\ref{fig:taxonomy} shows a preliminary taxonomy of the different model obfuscation methods and the best practice for model deployment. First, developers can use the \textit{renaming} and \textit{parameter encapsulation} to prevent most model parsing or reverse engineering tools from extracting the information of the deployed model. In the scenarios where computational costs are critical, developers can use the \textit{\thirdObfuscateStrategyName} and \textit{shortcut injection}, as they do not introduce any additional overhead. For structure obfuscation, developers can use \textit{shortcut injection} and \textit{extra layer injection}. These methods can significantly increase the difficulty of understanding the model structure of the deployed models. In addition, developers can use \textit{renaming} and \textit{\thirdObfuscateStrategyName} to disguise the deployed model, which can mislead the attacker into choosing the wrong architecture to produce a surrogate model. If the size of the app package is critical (e.g., deploying the model on devices with limited storage), developers need to carefully consider the trade-off between the number of obfuscated layers leveraging \textit{extra layer injection} with \textit{renaming}. The reason is that if \textit{renaming} is used to create different obfuscated layers for each layer, \toolname needs to create the corresponding APIs in the library to support obfuscated layers, hence increasing the library size.
% \vspace{-1em}
\subsection{\toolname \vs White-box Gradient Attacks}
In Section \ref{RQ4}, we show that our method can prevent existing reverse engineering methods from extracting the information of on-device models. However, what will happen if attackers directly perform white-box gradient attacks on the obfuscated model? In this subsection, we investigate the attacking performance of white-box gradient attacks on obfuscated on-device models, and analyze the factor that disables such kinds of attacks. We use three well-known white-box gradient attack methods FGSM~\cite{goodfellow6572explaining}, BIM~\cite{kurakin2018adversarial}, and PGD~\cite{madry2018towards} to evaluate the resilience of our method to this kind of attack. We use the Foolbox tool~\cite{rauber2017foolbox} to generate these attacks. The results are shown in Table~\ref{tb:wbg}. We cannot directly produce attacks for on-device models because the TFLite model format does not support gradient computing. But we can first convert the TFLite model to the PyTorch model using model conversion tools. The on-device models without obfuscation can be successfully converted to the PyTorch model, and the white-box gradient attack methods can successfully generate attacks for the converted PyTorch model. However, as we mentioned in Section \ref{RQ4}, the model conversion tools cannot transform obfuscated models into PyTorch models. When we directly use the obfuscated models as the input of the attack-generating functions, the attack methods cannot generate attacks for the obfuscated model. This is because (1) first, the white-box attack methods cannot parse the information of the obfuscated model, and (2) second, they cannot compute the gradient information of on-device models. Thus, we can conclude our method can prevent attackers from generating white-box gradient attacks using existing tools (\eg model conversion tools and attack-generating tools). 

\begin{table}[t]
% \small
% \vspace{-1.0em}
\centering
\caption{Evaluation using white-box gradient attacks. `$\surd$': the attack method cannot generate attacks. `{\scriptsize \XSolid}': the attack method can generate attacks.}
\begin{tabular}{l|ccc}
\toprule
 \multicolumn{1}{c}{}  & FGSM & BIM & PGD \\
\hline
Without obfuscation  & {\scriptsize \XSolid} & {\scriptsize \XSolid} & {\scriptsize \XSolid}        \\
\toolname & $\surd$ & $\surd$ & $\surd$ \\

% (0, 20)  & - & - & -     & -   & -      & -    &  - &  -       & -    & -  & \textbf{-}        \\
\bottomrule
\end{tabular}
\label{tb:wbg}
\end{table}
% \vspace{-1em}
\subsection{How to Parse Obfuscated Models?} 
When ModelObfuscator obfuscates the model, it will create a cache file to guide the tool to generate a compatible DL library. Attackers cannot automatically extract the obfuscated model unless they obtain the cache file from the developer's computer (which is unlikely to happen).
% \xiao{Is this the only way to extract obfuscated models? Is it possible for attackers to manually reverse engineer (with significant manual efforts) the model if they have the model file and library file?}
Generally speaking, attackers must use reverse engineering to get source codes from the compiled library, which consists of binary files, and it will cost significant manual effort to understand the obfuscated model.

\subsection{Limitations}
Although the proposed model obfuscation does not introduce significant computational overhead, it will increase the size of the modified TFLite library. This is because we need to provide support for the new obfuscations made. For a huge network like a 1000-layer network deployed model (although it is unlikely to find a such deployed model in the real world), the size of the modified TFLite library will significantly increase if we rename every layer. As a result, the app package also increases as the modified TFLite library will also need to be deployed on the device. Besides, our method cannot defend against query-based attacks as they do not need the inner information of models. But they are less effective than white-box attacks, which can be disabled by the proposed \toolname.

% \xiao{I think it is unlikely for the attacker to obtain files on the developer's computer. I suggest not to mention this as a limitation, or to say it is unlikely to happen.}
% Avoiding using the copy mechanism for renames and hence reducing memory overhead is left as future work.
\section{Conclusion}
% \vspace{-0.5em}
In this work, we analyzed the risk of deep learning models deployed on mobile devices. Attackers can extract information from the deployed model to perform white-box-like attacks and steal its intellectual property. To this end, we proposed a model obfuscation framework to secure the deployed DL models. We utilized five obfuscation strategies to obfuscate the information of deployed models, \ie \textit{renaming}, \textit{parameter encapsulation}, \textit{neural structure obfuscation}, \textit{shortcut} \& \textit{extra layer injection}. We developed a prototype tool \toolname to automatically obfuscate the TFLite model and produce a compatible library with the model.
Experiments show that our method is effective in resisting the model parsing tools without performance sacrifice. Although the \toolname will increase the library size, considering the acceptable time \& memory overhead required and the extra security it achieves, we believe our method has the potential to be an essential step for security-sensitive smart apps. In future works, optimizing model obfuscation to reduce the library size is worthwhile to be explored.
\begin{acks}
Zhou is supported by a Monash Graduate scholarship.
Grundy is supported by ARC Laureate Fellowship FL190100035. Gao is supported by the National Natural Science Foundation of China under Grant No (62202026). Also, our sincere gratitude goes to Jiawei Wang, a PhD student at Monash, for providing invaluable help in solving problems of TFLite modification.
\end{acks}

\bibliographystyle{ACM-Reference-Format}
\bibliography{acmart}

\end{document}